\documentclass[
    ,final            
  ]
  {aipproc}

\layoutstyle{8x11double}

\usepackage{graphicx}
\newcommand{\abb}[1]{Fig.\,\ref{#1}}
\newcommand{\msun}{\ensuremath{\, {\rm M}_\odot}}
\newcommand{\cdr}{\ensuremath{^{13}\mem{C}}}
\newcommand{\kelv}{\ensuremath{\,\mathrm K}}
\newcommand{\lsun}{\ensuremath{\, {\rm L}_\odot}}
\newcommand{\nezw}{\ensuremath{^{22}\mem{Ne}}}
\newcommand{\ndr}{\ensuremath{^{13}\mem{N}}}
\newcommand{\czw}{\ensuremath{^{12}\mem{C}}}
\newcommand{\nvi}{\ensuremath{^{14}\mem{N}}}
\newcommand{\mem}[1]{\ensuremath{\mathrm{ #1}}}
\newcommand{\jahre}{\ensuremath{\, \mathrm{yr}}}

\begin{document}

\title{Nuclear burning and mixing in the first stars: entrainment at a convective boundary using the PPB advection scheme}

\classification{95.30.Lz, 95.30.Tg, 95.75.Pq, 97.10.Cv, 97.20.Li, 97.20.Wt}
\keywords      {first stars: nucleosynthesis, mixing, hydrodynamics}

\author{Paul Woodward}{
  address={Laboratory for Computational Science \& Engineering, University of Minnesota, USA}
}

\author{Falk Herwig}{
  address={Astrophysics Group, School of Physical and Geographical Sciences, Keele University, UK},
altaddress={T-Division, Los Alamos National Laboratory, Los Alamas, NM, USA (Affiliate)}
}

\author{David Porter}{
  address={Laboratory for Computational Science \& Engineering, University of Minnesota, USA}
} 

\author{Tyler Fuchs}{
  address={Laboratory for Computational Science \& Engineering, University of Minnesota, USA}
} 

\author{Anthony Nowatzki}{
  address={Laboratory for Computational Science \& Engineering, University of Minnesota, USA}
} 

\author{Marco Pignatari}{
  address={Astrophysics Group, School of Physical and Geographical Sciences, Keele University, UK}
}

\begin{abstract}
  The evolution of the first generations of stars at zero or extremly
  low metallicity, and especially some crucial properties like the
  primary \nvi\ production, is charactarized by convective-reactive
  mixing events that are mostly absent from similar evolution phases
  at solar-like metallicity. These episodes occur when unprocessed
  H-rich material is mixed accross a convective boundary into \czw\
  rich He-burning material, as for example in He-shell flashes of
  extremely-low metallicity AGB stars. In this paper we describe the
  astrophysical context of such convective-reactive events, including
  the difficulty of current one-dimensional stellar evolution models to
  correctly simulate these evolutionary phases. We then describe the
  requirements and current state of modeling convective-reactive
  processes in the first stars environment. We demonstrate some of the
  new concepts that we are applying to this problem, i.e. the highly
  accurate PPB advection scheme in the framework of PPM hydrodynamic
  simulations of mixing accross a very stiff convective boundary. We
  show initial results of such simulations that address the first
  non-reactive step of this problem, which is the entrainment of H at the top
  boundary of the He-shell flash convection zone.
\end{abstract}

\maketitle


\section{Introduction}
The formation and evolution of the first generations of stars traces
the formation and evolution of the first structure on cosmological
scales (e.g.\ O'Shea, this volume), and starts the continuing process
of chemical evolution in galaxies. The nuclear production in these
stars provides an important tracer of events that took place in the
early Universe \citep[e.g.][]{frebel:07}. An increasing number of very
metal-poor stars ($\mem{[Fe/H]} < -2$) are now discovered in large
spectroscopic surveys \citep{beers:05} reaching down to metalicities
below $\mem{[Fe/H]}=-5$ \citep{aoki:06a}. These low-mass stars
preserve the signature of the nuclear processes in the first
generations of stars, of both low mass as well as from massive
stars. Some important findings from the observational study of stars
that carry the signature of the first generations of stars include the
following.

The fraction of C-enhanced stars ($\mem{[C/Fe]}>1$ ) is between about
$10\%$ \citep{cohen:05} and $20\%$ \citep{lucatello:06} for stars with
$\mem{[Fe/H]} < -2$, and increases with decreasing metallicity. Below
$\mem{[Fe/H]} < -3$ there are now about $300$ C-rich stars, below
$\mem{[Fe/H]} < -3.5$ $40\%$ are C-rich. $80\%$ of these CEMP stars
are enhanced with s-process elements \citep{aoki:07,beers:07a}. Many
of these CEMP-s stars show evidence for binarity and
\citet{lucatello:05} find that present data is statistically
consistent with all CEMP-s stars beeing in binaries. The abundance
signature of this large group of CEMP-s stars has been interpreted as
beeing the result of mass transfer from an Asymptotic Giant Branch
(AGB) star, that is now a white dwarf. An additional important finding
is that CEMP-s stars have C to N ratios that are slightly lower than
predicted by non hot-bottom burning, low-mass, third dredge-up AGB
stars ($0 \leq \mem{[C/N]} \leq 1.5$), but about 2 orders of magnitude
larger than any hot-bottom burning AGB models \citep{johnson:07}.

As far as heavy elements are concerned one of the most remarkable
findings is the discovery of the solar-scaled, apparently universal
r-process element distribution for for the heavy neutron capture
elements, $56\leq \mem{Z} < 83$ \citep{cowan:06}. However, for lower
elemental mass down to Sr the observed element abundance distribution
in several extremely metal poor stars can not be only accounted for by
the standard solar-scaled r-process. Since at very low metallicity the
average contribution from the weak and main s process is negligable,
an extra contribution from a primary s process called light element
primary process (LEPP) has been proposed
\citep{travaglio:04}. Recently, \citet{montes:07} showed that the LEPP
signature at very low metallicity could be reproduced by a process
with neutron densities of less than $10^{13} \mem{cm^{-3}}$, which are
intermediate between the typical s- (much lower) and r-process (much
higher). Concerning the light elements, the observations of extremely
metal poor stars of any variety point at a primary production of \nvi\
that is currently not yet well understood \citep[e.g.][]{chiappini:06}.

These observational findings of the first generations of stars, that
have been discussed at length at this First Stars III conference, have
one thing in common. They are sensitively related to the physics of
mixing, and in some cases the hydrodynamic interaction of mixing and
nuclear burning. The present 1-D spherically symmetric stellar
evolution models of these low-metallicity stars do not account
properly for this physics. While this is true also for solar-like
metallicity, the uncertainties are amplified by the flash-like burning
of H mixed into \czw-rich He-burning layers. As we will describe below
these convective-reactive events are peculiar to extremely
low-metallicity stellar conditions. Uncertainty in mixing is not
limited to convection-induced mixing of course. Rotation must play an
important role. The uncertainty of present rotating stellar models is
evident from their inability to properly meet constraints from
pulsars \citep{heger:00,hirschi:05} in the case of massive stars
models, or the inhibiting effect of rotation in simulations of the s
process in low-mass stars \citep{herwig:02a}. 
However, some areas related to the comparatively fast
and dynamic evolution of convective and convective-reactive events in
the stellar interior are open to detailed hydrodynamic simulations. In
the following sections
we will discuss the discussed observational properties of the
extremely metal poor stars in the framework of one-dimensional stellar
evolution simulations with the goal to identify the most promising,
high-gain situations in first stars that can be investigated by
hydrodynamical simulations now. We will discuss the one-dimensional
spherically symmetric assumption in the convective-reactive cases. We
will then present some new simulations of one of these important
convective boundaries, showcasing the PPB advection scheme at work
inside the most recent version of the LCSE PPM stellar hydodynamics
code. Finally we report high-resolution simulations of the top
convective boundary of He-shell flash convection. Mixing at this
boundary determines the properties if the H-ingestion flash in EMP AGB
stars.

\section{Stellar evolution at zero or extremely low metallicity}
\label{sec:evol}

\subsection{Hot-bottom burning and hot dredge-up}

In the Introduction we have summarized the observational properties of
the CEMP-s stars. Low-mass AGB stars below $2\msun$ are indeed
expected to produce C- and s-rich abundance signatures even at
extremely low metallicity \citep[e.g.][]{travaglio:01a}. But slightly more
massive primaries would have experienced the third dredge-up
\emph{and} hot-bottom burning with a characteristically low C/N
ratio in the ejecta. According to Campbell's grid calculation
\citep{campbell:07a,campbell:07b} at an iron content of
$\mem{[Fe/H]}\leq -3$ all models with $M\geq 2\msun$ show at the surface the CNO
equilibrium value $\mem{[C/N]} \sim -2$ because of
hot-bottom burning. From this it follows that if CEMP stars result from AGB binary mass transfer we should expect also a
certain number of N-enhanced metal poor (NEMP)
stars. \citet{johnson:07} have searched for these NEMP stars, looking
especially into possible selection effects and observable biases. They
did not find a single NEMP star in their targeted
investigation. Uncertainties in the treatement of mixing in
one-dimensional models rather exacerbate the problem as they include
possibilities to produce N even more efficiently at lower metallicity.

At face value standard stellar evolution predicts that the
nucleosynthetic source of the abundance signature in CEMP-s stars (or
any CEMP star thought to be poluted by a former AGB companion with
$\mem{[C/N]} > 0.0$) can only come from stars with initial masses less
than 2\msun. The exact value of this limiting mass including the
metallicity dependence still needs to be determined. It depends on the
physics of mixing in at least two ways. One is through the mixing
length parameter used in the convective envelope. Most standard
stellar evolution calculations use the solar scaled mixing length
parameter also during the AGB phase. In their text book
\citet[][Ch.\,14.3]{cox:68} recommend a mixing length parameter of
$1.5$ for a solar-type convection zone, and about $2.1$ for a
red-giant type convection zone. A larger mixing-length parameter for
AGB stars is also resulting from full hydrodynamic simulations by
\citet[][$\alpha_\mem{MLT}=2.7$]{porter:00} as well as the majority of
semi-empirical, observations based determinations \citep[][
$\alpha_\mem{MLT} = 2.2 \dots 2.6$]{mcsaveney:07}\footnote{In both
  cases mixing-length parameters were determined in the MLT version of
  \citet{cox:68}.}. A larger mixing-length parameter leads to more
vigorous envelope burning, and further decreases the limiting initial
mass of envelope burning. An update of the work by \citet{porter:00}
for the low masses and metallicities relevant here, and the
application of these results to one-dimensional stellar evolution
calculations would greatly enhance the accuracy of the determination
of the limiting mass for hot-bottom burning in EMP AGB stars.

The other mixing process that affects N production in low-mass AGB
stars is any mixing at or accross the bottom boundary of the
convective envelope. The slightly reduced observational [C/N] ratios
compared to low-mass predictions may hint at some mixing processes
below the convective envelope. Clearly, the whole radiative
\cdr-pocket based s-process concept \citep{gallino:97b} is built on
mixing of H accross the convective boundary into the \czw-rich
core. Whatever the physical cause is of these convection-induced
extra-mixing processes, it may easily enhance the hot-bottom burning
efficiency. This question can be investigated with some numerical
experiments similar to the one we present in the next section.

Modeling the production of C, N and s-process element production in
EMP AGB stars poses more challenges related to the mixing during the
third dredge-up. As we just mentioned, this mixing plays an essential
role in the formation of the \cdr\ pocket. What are the properties of
this mixing in EMP AGB stars?  Recent studies
\citep{herwig:03a,herwig:03c} have shown that even small amounts of
mixing at the bottom boundary of the convective envelope in
intermediate mass stars ($4$ and $5\msun$) can lead to possibly
violent, even eroding or excavating, flame-like burning.

\begin{figure}
 \includegraphics[height=.295\textheight]{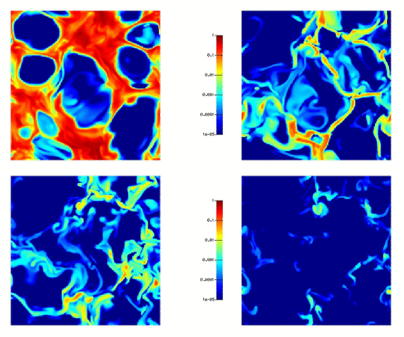}
 \caption{\label{fig:sbm05-1frame} A snapshot of the abundance of
   H-rich material (red, originally only in the radiative layer) in a
   HIF toy simulation. The panels show horizontal planes (from left to
   right and top to botom: just above the top convective boundary, and
   at three successively deeper location inside the convectively
   unstable layer) from an underresolved (150$^2 \times$100)
   three-dimensional RAGE simulation of He-shell flash convection. The
   setup and stratification is similar to \citet{herwig:06a}, with a
   much enhanced energy driving for the convection. Convective
   boundaries are not resolved, nevertheless the simulation
   demonstrates one of the reasons why convective-reactive events need
   to be investigated in the multi-dimensional hydrodynamic rather
   than one-dimensional, sphercially symmetric framework, as discussed
   in the text. }
\end{figure}
Fig.\,4 in \citet{goriely:04} demonstrates how the temperature at the
base of the convective envelope increases with mass and decreasing
metallicity, exceeding at $\mem{[Fe/H]}=-2.3$ $6 \cdot 10^7\kelv$ for
$4\msun$ and $8 \cdot 10^7\kelv$ for $5\msun$. In that work it is
described how the amount of mixing accross the convective boundary
that is associated with the formation of a \cdr\ pocket for the s
process at solar metallicity does not lead to a formation of s-process
elements at extremely low metallcity, due to instantaneous burning of
protons in the \emph{hot} third dredge-up of a 3\msun, Z=0.0001
case. \citet[][Fig.\,7]{herwig:03c} shows how this hot dredge-up can
feed back into the stellar structure and explain how this mixing,
depending on its efficiency leads to very deep, eroding
dredge-up. Mixing at this boundary with only a quarter of the
efficiency that is deduced from s-process observations for low-mass,
solar metalcity stars \citep{lugaro:02a} would within $2000\jahre$
completely eject the entire envelope through a high-luminosity ($\log
L/\lsun \sim 5$) driven superwind with mass loss rates of a few
$10^{-3}\msun/\jahre$. The luminosity during the hot dredge-up and the
consequences for the structural evolution and s process depend
sensitively on the mixing efficiency at the convective boundary. This
mixing efficiency is the result of a hydrodynamic convective-reactive
phase in which local mixing of the H and \czw\ interacts with nuclear
burning. This is locally an inherently three-dimensional process that
requires calibrated mixing models to be included into one-dimensional
simulations. The exact nature of this hot dredge-up mixing will
critically determine the subsequent evolution of EMP AGB
stars. Similar to the convective boundary simulations presented in the
next section, numerical hydrodynamic experiments to study the nature
of hot dredge-up mixing in EMP AGB stars are now within reach, and we
plan to carry those out in the near future. These simulations will
provide valuable insight into the s-process in EMP AGB stars and for
the fate of intermediate mass EMP stars during the TP-AGB phase. The
hot dredge-up may be related to the observational paucity of NEMP
stars.

\subsection{H-ingestion flashes (HIF)}
From what we have said so far about the s process it is clear that
there is presently considerable uncertainty about where and how the
heavy slow neutron-capture elements are made in EMP AGB stars. Future
stellar evolution calculations have to clarify if the temperature at
the bottom of the convective envelope during the third dredge-up is
low enough in the lowest mass AGB stars so that the \cdr-pocket for
the radiative s-process can form. It also remains to be seen what the role of the
convective \nezw\ neutron source is in the low-mass EMP AGB stars, as
it is of primary nature. Its effect although marginal in solar-like
metallicity AGB stars may be important at these extremely low
metallicities, possibly even allowing the primary formation of Fe
seeds for heavy-element production \citep{husti:07}.

However, there is another important alternative, or additional
production site for the s process in EMP AGB stars to be
investigated. Several authors have described the H-ingestion flash
(HIF) in thermal pulse stars of extremely low metallicity
\citep[e.g.][]{cassisi:96,fujimoto:00,chieffi:01,siess:02,herwig:03a}. The
HIF is related to the hot dredge-up described above. While in the
latter situation protons are mixed from the convectively unstable
envelope into the radiative \czw-rich core, the HIF relates to the
situation when protons are entrained from a radiative layer into the
He-burning convection zone, either during the He-core flash on the tip
of the RGB \citep[e.g.][]{hollowell:90b,schlattl:01}, the He-shell
flash in thermal pulse AGB or even in some massive star models of zero
metallicity during the He-core burning. HIFs occur at extremely low
metallicity because of a minimal entropy barrier between the
He-burning and hot H-burning layers. The large temperature of the
latter is a result of the low abundance of CNO catalytic material for
H-burning. HIFs are also known to happen in solar-like metallicity
environments. Post-AGB stars that have evolved through the
central-star of planetary nebulae phase may experience the HIF as a
very-late thermal pulse as young white dwarfs \citep[][andreference
therein]{werner:06}. Observed stellar matches are the born-again
stars, as for example Sakurai's object \citep{vanHoof:07}. Other
examples include accreting white dwarf that may be SN Ia progenitors
\citep{cassisi:98}, the post-RGB late He-flashers \citep{brown:01} or
even the X-ray burst phenomenon \citep{piro:07}. However, the HIF may
be especially important in the EMP AGB star environment as it provides
a neutron rich environment which may be playing the possibly dominant
role in the s-process nucleosynthesis in the most metal-poor AGB
stars.  \citet{iwamoto:04}, building on concepts layed out previously
by \citet{malaney:86}, has tentatively analysed this s-process
environment with interesting observable consequences.

\subsection{HIFs and hot dredge-up in massive stars}

Convective-reactive events very similar to the HIF and hot dredge-up
discussed so far for AGB stars are encountered in
stellar evolution models of zero or extremely low metallicity massive
stars \citep{limongi:01}.  After the
end of He-core burning the convective instability of the 
extended H-burning shell 
spreads inward in Lagrangian mass coordinate. Eventually the bottom of
this convective zone at which vigorous H-shell burning takes place
will be in immediate contact with the radiative layer underneath in
which He-burning has produced a small amount of primary \czw. This
amount of \czw\ is sufficient to instantaneously enhance the effective
rate of H-burning if brought into contact with the burning
shell. 

In stellar evolution calculations great care has to be taken to
numerically resolve the convective boundary. Instantaneous mixing
inside the convection zone without any mixing model for the boundary
is not resolvable in one-dimensional stellar evolution calculations,
and may easily lead to erratic and unphysical model results. Mixing
may be minimal at convective boundaries in massive star stellar
interiors, and formally it can be made as small as one can numerically
afford to resolve. However, if we take the interior convection
simulations of C- and O-shell burning of \citet{meakin:06a} as any
guidance, then such mixing at convective boundaries may not at all be
minimal.  The effect of this convective boundary mixing has already
been shown to be significant on the subsequent supernova explosion at
solar-like metallicity \citep{young:05b}, and it can be expected that
similar mixing at all convective boundaries with their respective
efficiencies will be even more significant at the low metal content
of the first stars.
\begin{figure}
 \includegraphics[height=.35\textheight]{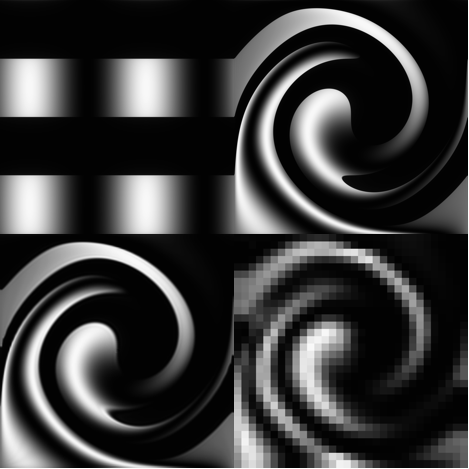}
 \caption{\label{fig:2DadvectionPPM} A 2-D advection test of the PPB
   scheme.  In the upper panels, the initial distribution (left) and
   final one (right) are shown on a $1024^2$ grid.  At the bottom left
   and right we show the final distributions on grids of $128^2$ and $16^2$,
   using the 6 moments updated by the scheme in each grid cell to
   generate average values in cell quadrants.}
\end{figure}

HIF-like events can also be encountered in massive first stars when the
He-core burning convection zone entrains H from the overlying radiative layers (S.\ Ekstr\"om, priv.\ com.).  HIF and hot
dredge-up type convective-reactive events can be initiated by
rotationally induced mixing \citep{hirschi:07}, and are routinely
found in such stellar models. These convective-reactive phases may be
especially relevant to nucleosynthesis in extremely metal-poor massive
stars, for example for investigating the LEPP (see Introduction) as
this mixing determines how much primary \cdr\ and \nvi\ is produced as
a neutron source for heavy element production.

\subsection{Why the need for three-dimensional simulations}
One-dimensional stellar evolution models of the HIF in AGB stars,
e.g.\ those by \citet{herwig:03a} or \citet{iwamoto:04}, are based on
two assumptions that are not obvisouly appropriate for
convective-reactive events. These are spherical symmetry and mixing as
a diffusion process. Convective mixing is not a diffusion but rather
an advection process.  The introduction of the diffusion formalism to
treat mixing, e.g.\ in \citet{herwig:97}, was motivated by the finding
of \citet{freytag:96} that overshoot mixing outside the formally
unstable convection zone behaves like diffusion. However, relying on
diffusion for the mixing framework in convective-reactive situation
inside the unstable region \citep[as for example in][]{herwig:03a} is
at least questionable. Spherical symmetry can be broken on a global
scale, as for example by very fast rotation, and this is why 1D
stellar evolution often assumes in such cases a shellular rotation law
\citep{meynet:97}. One may assume that a HIF is still well described
by global spherical symmetry, although even that is not
clear. H-deficient high-velocity ($200\mem{km/s}$) knots are moving
away from the H-deficient planetary nebula central stars Abell 30 and
Abell 78 \citep{borkowski:93} in an otherwise H-normal PN. These two
stars are believed to be post-very late thermal pulse stars (see
above), i.e. a HIF in a pre-white dwarf. If the knots are ejections
that originate HIF hydrodynamics then this could imply that global
spherical symmetry is violated in the HIF. On a local scale spherical
symmetry is questionable because the assumption that the H-abundance
is exactly homogeneous on shells has not been shown to be valid. The
answer to this question lies in the competing effect of small scale
turbulent mixing that enforces homogeneity and the large-scale
convective motions that generate inhomogeneities. He-shell flash
simulations of \citet{herwig:06a}, or the 2D simulation discussed
below may indicate that convective downdrafts contain locally more
entrained H compared to updwelling material that originates at the
He-burning convection zone bottom. This difference leads to a
patchiness of the H abundance in any given horizontal plane, resulting
in a patchiness of the energy generated from proton captures by \czw.
\abb{fig:sbm05-1frame} demonstrates this effect through
inhomogeneities in horizontal planes of a toy model. This needs to be
studied quantitatively in properly resolved simulations with the aim
to investigate how thick the H-burning layer is, and how the burning
products \ndr\ and \cdr\ will be distributed in the He-intershell.
For the same reasons we need to extend hydrodynamic simulations also
to the hot dredge-up episodes, in both intermediate and massive stars.

\begin{figure}
  \includegraphics[height=.177\textheight]{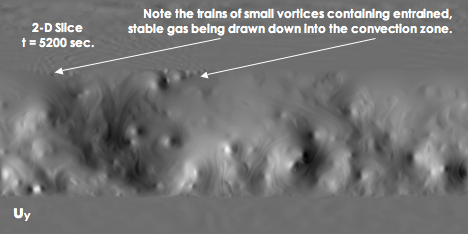}
  \caption{\label{fig:2DPPMHe-shellflashconv} A 2-D simulation of
    turbulent convection in the helium shell flash convection zone of
    a two-solar-mass AGB star.  The initial 1-D base state about which
    this PPM code computes the perturbation resulting from simulated
    helium shell flash energy input approximates closely the 1-D
    stratification from the stellar evolution simulation.  In this
    image the vertical component of the velocity is displayed.  This
    $1024 \times 512$ grid reveals trains of small vortices that form at the
    top of the convection zone where the gas descends.  This is the
    region and the characteristic flow pattern that we have used to
    initialize our 3-D PPM simulations reported in the next section.}
\end{figure}

\section{Stellar interior hydrodynamics simulations}
\label{sec:hydro}

A first step toward comprehensive simulations of the HIF are simulations of the top boundary of
the He-shell flash convection zone where the proton-rich material will
be entrained. The exact nature of mixing at this boundary determines
the rate of H-ingestion as the convective shell makes first contact
with the H-rich radiative layer above.  

\subsection{The numerical technique}
\subsubsection{A fully compressible, explicit PPM code for low-Mach number flows}

Our simulations are carried out with an explicit gas dynamics code,
multifluid PPM, or PPMMF, which accurately tracks sound wave signals
even at the low Mach numbers of about up to 1/30 that are found in
convective gusts. We do this of course, because these are the tools we
have at hand.  However, a non-directionally-split gas dynamics
algorithm must involve roughly 3 times as much work as a directionally
split one if it is to achieve the same accuracy.  If such a method is
made implicit, and only such methods can be made implicit, then an
additional cost per time step of at least a factor of 2 results.  This
means that explicit codes are cost competitive down to at least Mach
1/6, and probably down to Mach 1/10 as a practical rule of thumb.  Our
PPM code has a special implementation in terms of 1-D passes using
short vector lengths of 16 exclusively on 128-byte quadword operands,
which delivers very high levels of performance, exceeding one useful
flop per clock tick per processor core with all costs of the code
included. It is doubtful that an implicit algorithm, without this 1-D
operator decomposition and computational intensiveness, could enjoy
this same level of performance.  We therefore believe that our
computational approach is cost effective, despite the large number of
time steps we need in these stellar hydrodynamics problems. 

Our version of PPM \citep[cf.][]{woodward:06b} modifies the monotonicity constraints in
the original PPM so that they are not invoked at all in smooth regions
of the flow.  This more careful treatment results in considerable
improvements in low Mach number flows, which do not contain the
shock-generated discontinuities that motivated the constraints in the
original scheme.  We also note that PPM , unlike methods that use
linear interpolation, does not decrease appreciably in accuracy with
decreasing Courant number. The reason for this is its elaborate,
high-order interpolation of the values at the edges of grid cells,
which assume a paramount importance in explicit computations of low
Mach number flows.  This insensitivity to Courant number is shared by
the multifluid tracking algorithm in PPM, PPB, which is described
below.

\begin{figure}
  \includegraphics[height=0.85\textheight]{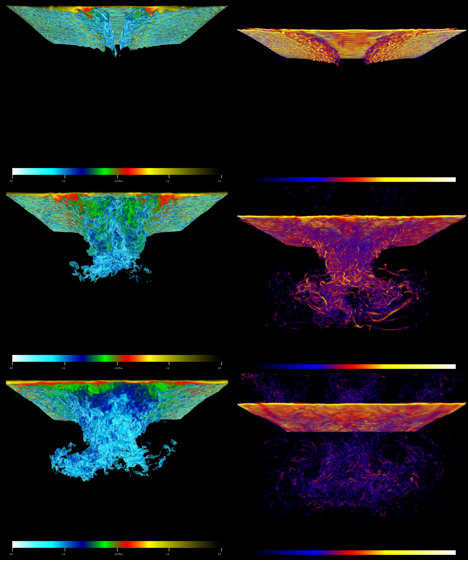}
  \caption{\label{fig:TopEntrain3DPPM} A 3-D simulation of entrainment
    of stably stratefied hydrogen and helium gas mxiture into the 2.26
    times denser helium and carbon mixture of the helium shell flash
    convection zone below it (see discussion in text).  The fractional
    volume images show mixing ratios by volume of 1 part in $10^4$ as
    aqua, 1 part in $10^3$ as blue, 1 part in 100 as red, and 1 part
    in ten or higher mixing fractions are yellow.  Cells containing
    only pure upper gas are transparent.}
\end{figure}

 \subsubsection{The PPB advection scheme}
 The relative concentrations by volume of the two fluids in our
 problem are tracked by the PPB advection scheme \citep{woodward:86},
 which is a properly constrained, 3-D extension of van Leer's 1-D,
 unconstrained Scheme VI \citep{vanleer:77}.  The scheme as described
 by \citet{woodward:86} has been substantially improved, increased in
 efficiency, and much more carefully constrained over the years since
 that article, and it has been implemented as a module of the XRAGE
 production code at Los Alamos in 2004.  It has roughly triple the
 resolving power of PPM advection, due to its use of triple the amount
 of independent information in each 1-D pass.  This extra information
 comes in the form of low-order moments of the distribution of the
 fractional volume variable in each cell.  In 3-D, 10 such moments are
 carefully updated by the scheme using 1-D passes, like the host gas
 dynamics scheme, PPM.  These 10 moments serve to define a parabola
 along any line within the cell for the fractional volume variable
 (they define a 10-coefficient quadratic form).  The updating of the
 moments is by conservation laws, which results in powerful special
 properties for the scheme \citep[see][]{woodward:86}.  We apply the
 PPB scheme to fractional volume advection here, and as a result we
 take very special care to handle the constraints applied to the
 implied quadratic form in the cell so that sudden transitions of this
 variable between the special values of 1 and 0 are properly
 represented.  These constraints are far less dramatic than those
 involved in PPM's contact discontinuity detection and steepening
 algorithm, and they deliver the same one-cell effective thickness of
 a fluid edge without anything like the side effects that contact
 discontinuity steepening can cause.  We have observed these benefits
 in numerous tests over many years.  Here we give only a single
 representative example.

 In \abb{fig:2DadvectionPPM} we show the results of 2-D advection in a
 swirling flow field similar to that encountered in our stellar
 hydrodynamics problem.  The initial distribution has smooth sine wave
 components in one dimension and relatively sharp, but not
 discontinuous components in the other.  The initial distribution is a
 product of a sine wave in x and the 16th root of the absolute value
 of a sine wave in y, with the proper sign of the sine wave in y
 restored.  The variation is then scaled to go from 0 to 1.  The
 sudden jumps in this distribution, shown in the upper left-hand panel
 of \abb{fig:2DadvectionPPM} , do not go between the special values of
 0 and 1.  Their treatment therefore reveals the very small extent to
 which Gibbs phenomena tend to reduce the faithfulness of this
 scheme's representation of jumps that do not go between the special
 values that are so carefully treated.  The flow field vanishes at the
 center and at the edges of the square domain, and thus this advection
 problem is nonlinear.  The results at the initial and at a subsequent
 time are shown on grids of $1024 \times 1024$, $128 \times 128$, and
 $16 \times 16$ cells.  To generate these displays, we have used the
 moments of the distribution within the cells that are updated by the
 PPB scheme (in 2-D we use 6 moments) to generate average values of
 the distribution in the 4 quadrants of each grid cell.  In the
 displays, each such quadrant is represented by a constant value, so
 that the individual quadrants and the resolution of the grid can thus
 be immediately perceived.

\subsection{Numerical Experiments in 2D}

Numerical experiments in 2D motivate the initialization of a 3-D
experiment. We first briefly summarize 2-D simulations of the helium
shell flash convection zone of a two-solar-mass star about a month
before the peak of the helium shell luminosity and before the
convection zone has had a chance to expand very much in radius (this
is the same 1-D initial state used in the simulations reported in
[13]).  In \abb{fig:2DPPMHe-shellflashconv} we see a display of the
vertical component of velocity in our simulation with the PPM code.
Here we use a fine grid of $1024 \times 512$ uniform grid cells.  We
also deal with the several orders of magnitude variation with height
of the pressure and density by subtracting out from the governing
equations the hydrostatic, 1-D base state and computing only the
perturbation about it.  The bottom of the convection zone, which is
easy to make out in Figure 2, is at a radius of 9150 km, and the top,
also easily discernable, is at 15,200 km radius.  At this time, the
hydrogen-rich envelope begins at radius 17,500 km.  The Mach number of
this flow is, at its highest 0.035, which makes this explicit
computation relatively expensive, although this calculation was easily
performed on a single desktop PC. This 2-D flow shows the tendency for
gas from the stably stratified layer above to be entrained at those
places where large convection cells flow along the top of the
convection zone and begin to descend.  Trains of small vortices formed
due to shear instabilities are seen descending with the general flow
in these regions, potentially carrying a bit of the stably stratified
upper gas along inside them.  Because the low Mach number of this flow
introduces some computational expense in 3D, we chose to include in
our 3-D simulation domain only the portion of the convection zone
where we expect this entrainment to occur.  We use periodic boundary
conditions in the two horizontal dimensions, and we, admittedly
artificially, introduce reflecting walls at the top and bottom of the
domain.  We place the top of the convection zone at the middle height
in our problem domain, which then extends $100\mem{km}$ down into the
helium shell flash convection zone below and an equal distance above
the top of this convection zone.  In the convection zone, we introduce
two Mach 1/30, counter-rotating, horizontally oriented large vortices
with an initially approximately incompressible flow field and filling
the region between the top of the convection zone and the bottom of
our grid.  We also introduce perturbations near the top of the
convection zone in the vertical component of the velocity with
sinusoidal variation in modes 1, 3, and 5 in the x-direction along the
length of the convective rolls.  To stimulate shear instabilities in
this dimension, we also introduce a small shear in the x-component of
the velocity at the top of the convection zone.  We will see below
that due to the stable stratification above the top of the convection
zone, this shear in x-velocity has little effect in comparison to the
behavior induced by the convective rolls.  We take the fluid states
above and below the top of the convection zone from later in the 1-D
stellar evolution, when above this mid-plane in our domain we have
mainly hydrogen and helium while below it we have mainly helium and
carbon with a density jump of a factor of 2.26 across the midplane
(the top of the convection zone).  The gas everywhere has a gamma-law
equation of state with a gamma of 5/3.

\subsection{Results of a 3-D Numerical Experiment}
Visualizations of the fractional volume of the upper gas are shown at
the left in \abb{fig:TopEntrain3DPPM}, and visualizations of the
vorticity are shown at the right at the same times, although from a
different viewpoint.   These
simulations show the capability of this 3-D approach. Preliminary
quantitative analysis indicates a partially mixed boundary layer ($10
- 15\mem{km}$) in which the abundance of the H-rich mix from the top
stable layer drops in horizontal averages from unity to below $\sim
10^{-4}$ after $20\mem{s}$ simulated time, which is about 1/30 of the
convective turn-over time scale. Signigicant inhomogeneities are
present in the mix layer with the most enriched downdrafts having
maximum H-rich material abundances of $10^{-2} \dots 10^{-3}$ at depth
of $40$ to $50\mem{km}$ below the boundary. Although these results are
based on a set of three simulations with grids of $256^3$, $512^3$ and
$1024^3$ convergence tests have not been finalized.

These 3-D simulations are meant to study mixing due to the shear at
convective boundaries that is induced by horizontal motions driven by
convection. The vertical momentum component is zero in this
setup. However, a simulation that contains the full convection zone
perturbations of the boundary will also include vertical components on
larger scales, and in fact these perturbations excite the internal
gravity wave spectrum that is observed in the stable layers above and
below the convection. 

Thus, mixing as found in these simulations
applies to what are rather small scales in the full convection
simulations. Nevertheless, even with these assumptions we find that
notable amounts of material cross the convective boundaries.
Here, as a result of our initialization of this problem, we do not
observe substantial distortion of the upper boundary of the convection
zone by the motions below it.  Instead we entrain buoyant gas from the
region above this boundary by essentially scraping off bits of it and
mixing it into the lower fluid that is moving past.  This mixing
allows the flow to carry the more buoyant fluid down into the
convection zone.  When the entrained fluid is hydrogen rich, even if
it becomes only a trace constituent of the descending gas it can
release significant amounts of energy through nuclear burning.


\begin{theacknowledgments}
  The work reported here has been supported at the University of
  Minnesota by grant DE-FG02-03ER25569 from the MICS program of the
  DoE Office of Science, by NSF equipment grants CNS-0224424 and
  CNS-0421423, and by the Minnesota Supercomputing Institute, a part
  of the University of Minnesota's Digital Technology Center.  Funding
  for this work was also provided through EU Marie Curie grant
  MIRG-CT-2006-046520. FH is Affiliate of the Theoretical
  Astrophysics Group in T-Division at LANL, and acknowledges continued
  collaborative support. We would like to thank Raphael Hirschi for
  extremely useful discussions. Brian O'Shea and the other organizers
  of the FSIII meeting have to be thanked for their enourmous
  flexibility to deal with various last-minute issues we imposed on
  them.

\end{theacknowledgments}

\bibliographystyle{aipproc}   


\IfFileExists{\jobname.bbl}{}
 {\typeout{}
  \typeout{******************************************}
  \typeout{** Please run "bibtex \jobname" to optain}
  \typeout{** the bibliography and then re-run LaTeX}
  \typeout{** twice to fix the references!}
  \typeout{******************************************}
  \typeout{}
 }

\end{document}